\documentclass[11pt]{article}
\usepackage{graphicx}
\usepackage{amssymb}
\usepackage{epstopdf}

\title{Enclosed area distribution in percolation}
\author{Robert Ziff, University of Michigan, Ann Arbor}
\begin{document}
\maketitle

(Talk presented in the StatPhys22 conference in Bangalore, India,
July 5-9 2004.)

\bigskip
{\noindent \bf \large Abstract}

\bigskip
The number of two-dimensional
percolation clusters whose external hulls enclose
an area greater than $A$, in a system of area $\Omega$,
behaves at the critical point
as $C \Omega /A$ for large $A$, where $C = 1/8 \pi \sqrt{3}$.
Here we show that away from the critical
point this factor is multiplied by a scaling function that is asymptotically proportional
to a simple exponential $\exp(-A/A^*)$ where $A^*$
scales as $|p - p_c|^{-2 \nu}$.  The fit is better
than for Kunz and Souillard sub-critical scaling, 
which would predict the asymptotic behavior $\exp(-(A/A^*)^{2/D})$
where $D = 91/48$ is the fractal dimension.
\bigskip
{
\section{Percolation}
In the percolation model, one considers a disordered
system, constructed (for example) by taking a regular
lattice and diluting it by making only a fraction
$p$ of the sites or bonds conducting (occupied).
One is concerned with the appearance of long-range
connectivity in the system -- the percolation transition --
and the behavior near the point of that transition. 
For definiteness, we consider mainly
bond percolation on a square lattice, and define
\begin{equation}
s = \hbox{the number
of wetted sites of a cluster.}
\end{equation}
Isolated sites with no bonds attached
correspond to $s=1$.

\centerline {
\includegraphics[width=5 in]{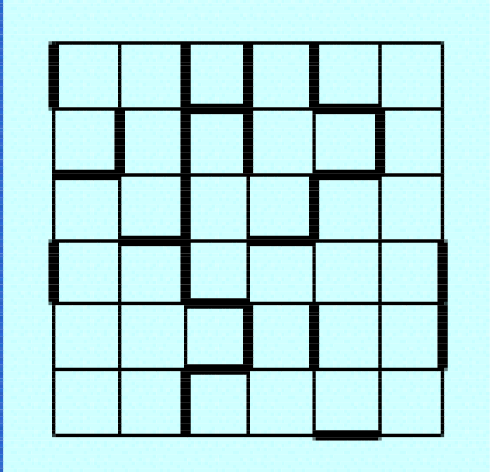}}

{\large Bond percolation on a square lattice, showing
the appearance of a percolating path from the
top of the system to the bottom}

Following are pictures of a single system
where $p$ is slowly increased (bonds are
added) and clusters are merged (as in the
Newman Ziff algorithm).

\centerline {
\includegraphics[width=5in]{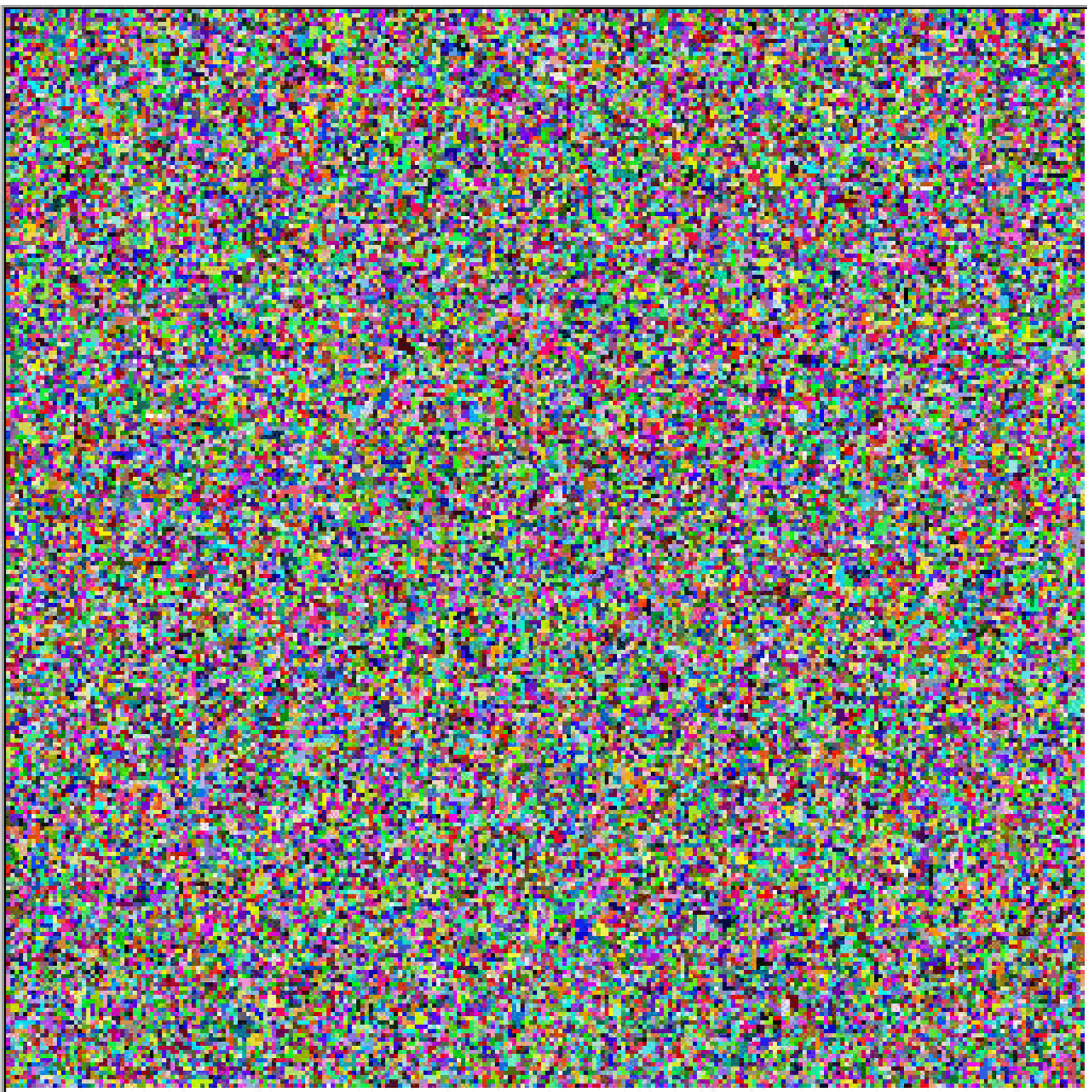}}

\centerline {
\includegraphics[width=5in]{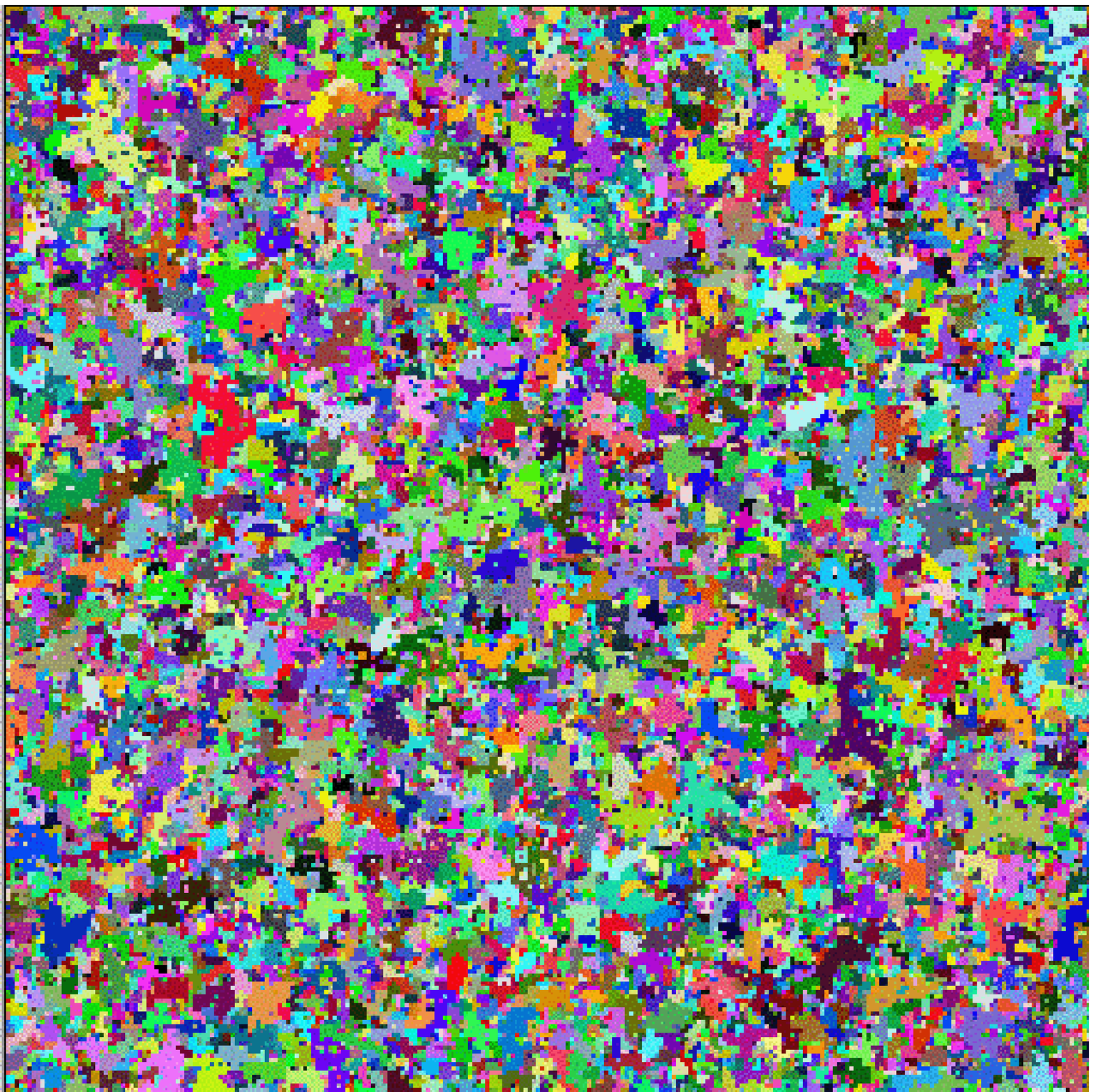}}

\centerline {
\includegraphics[width=5in]{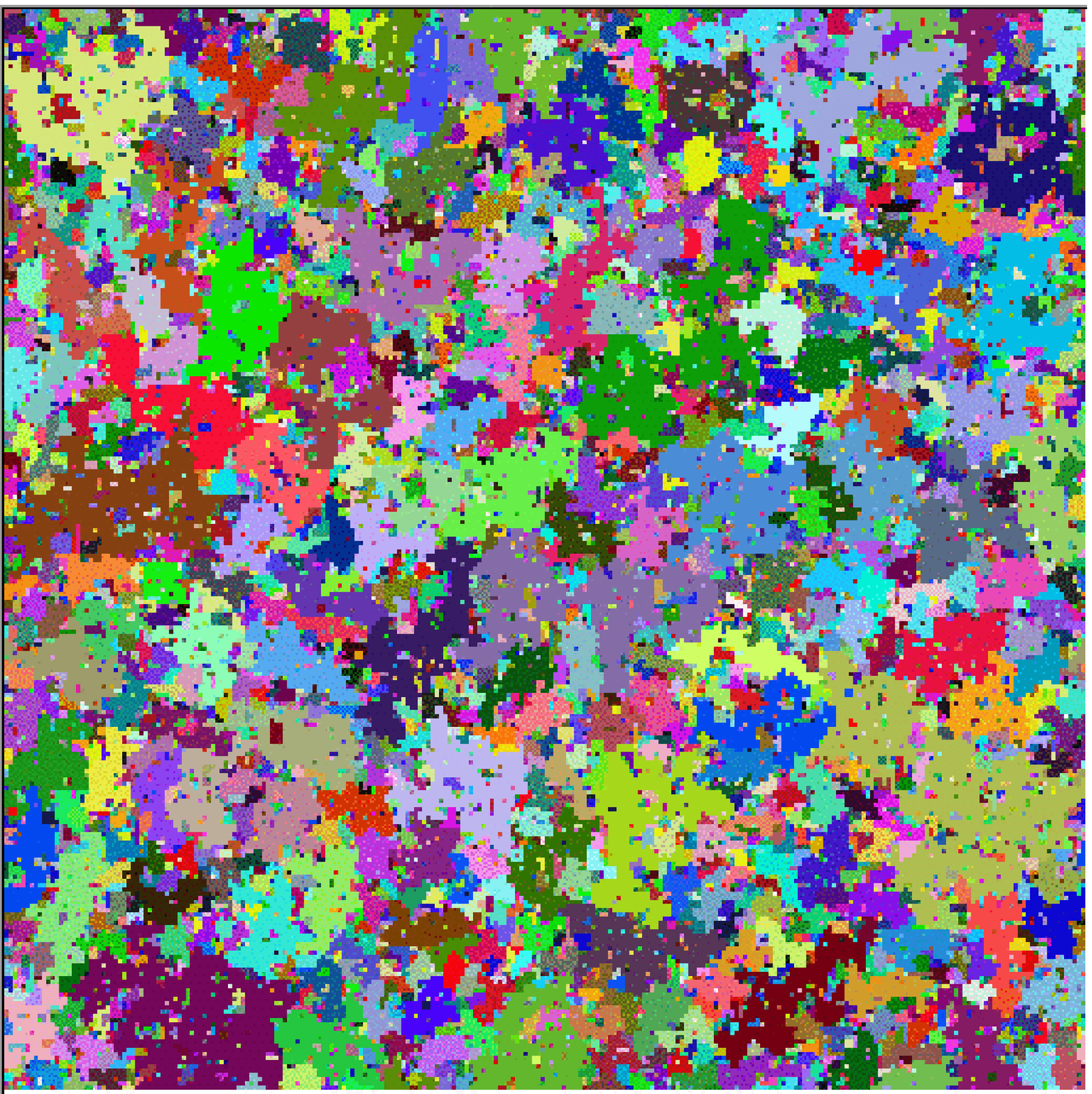}}

\centerline {
\includegraphics[width=5in]{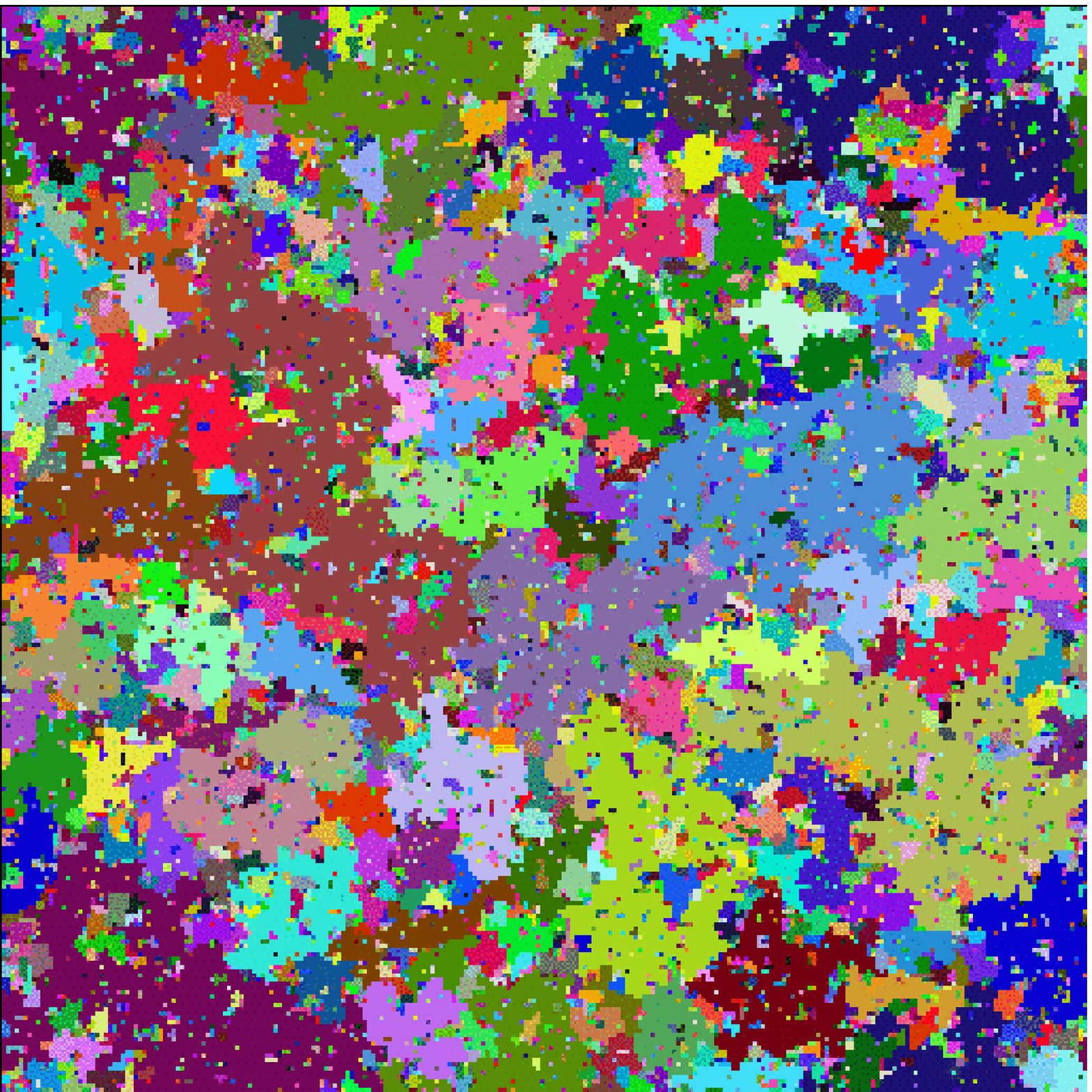}}

\centerline {
\includegraphics[width=5in]{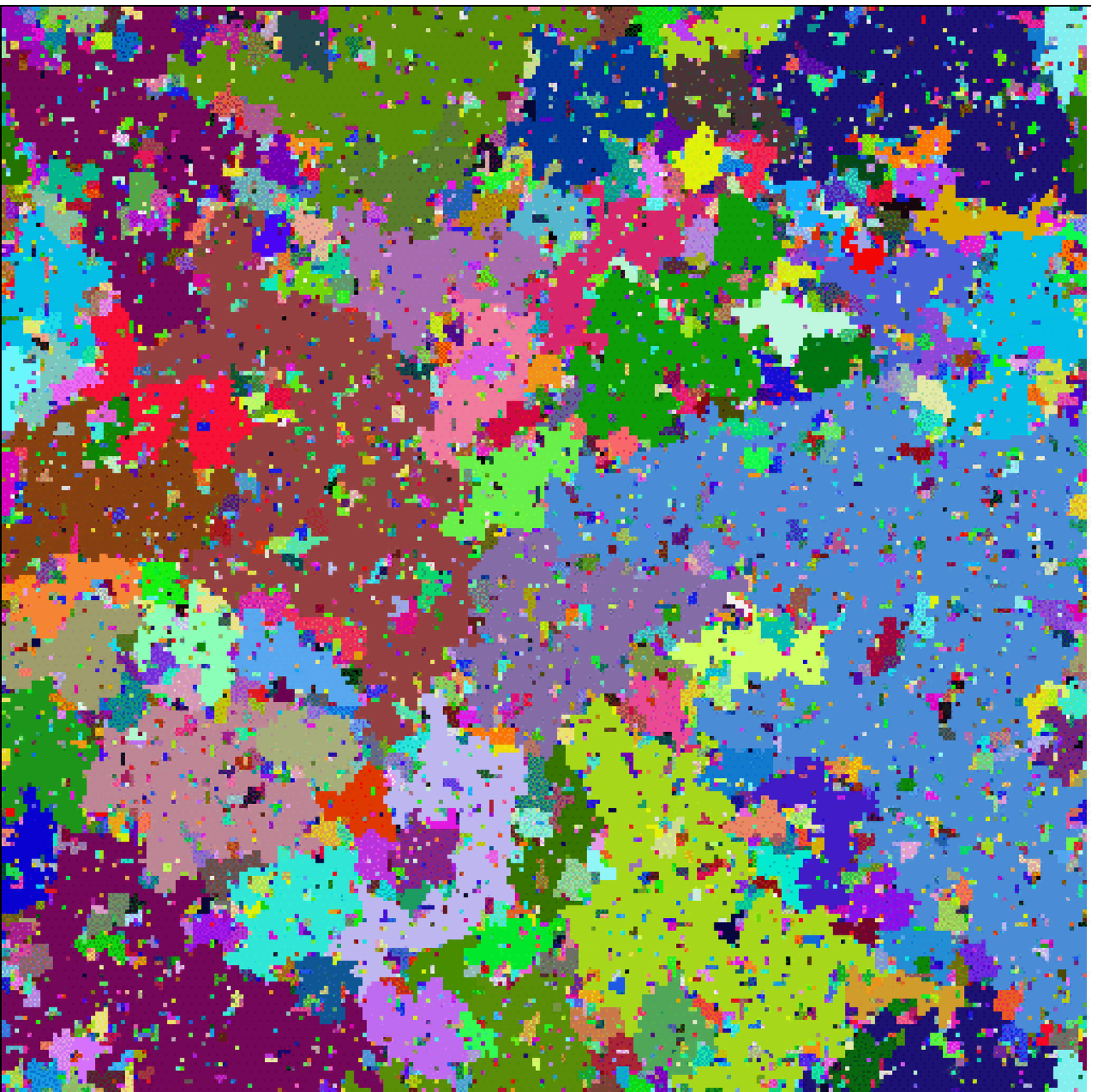}}

\centerline {
\includegraphics[width=5in]{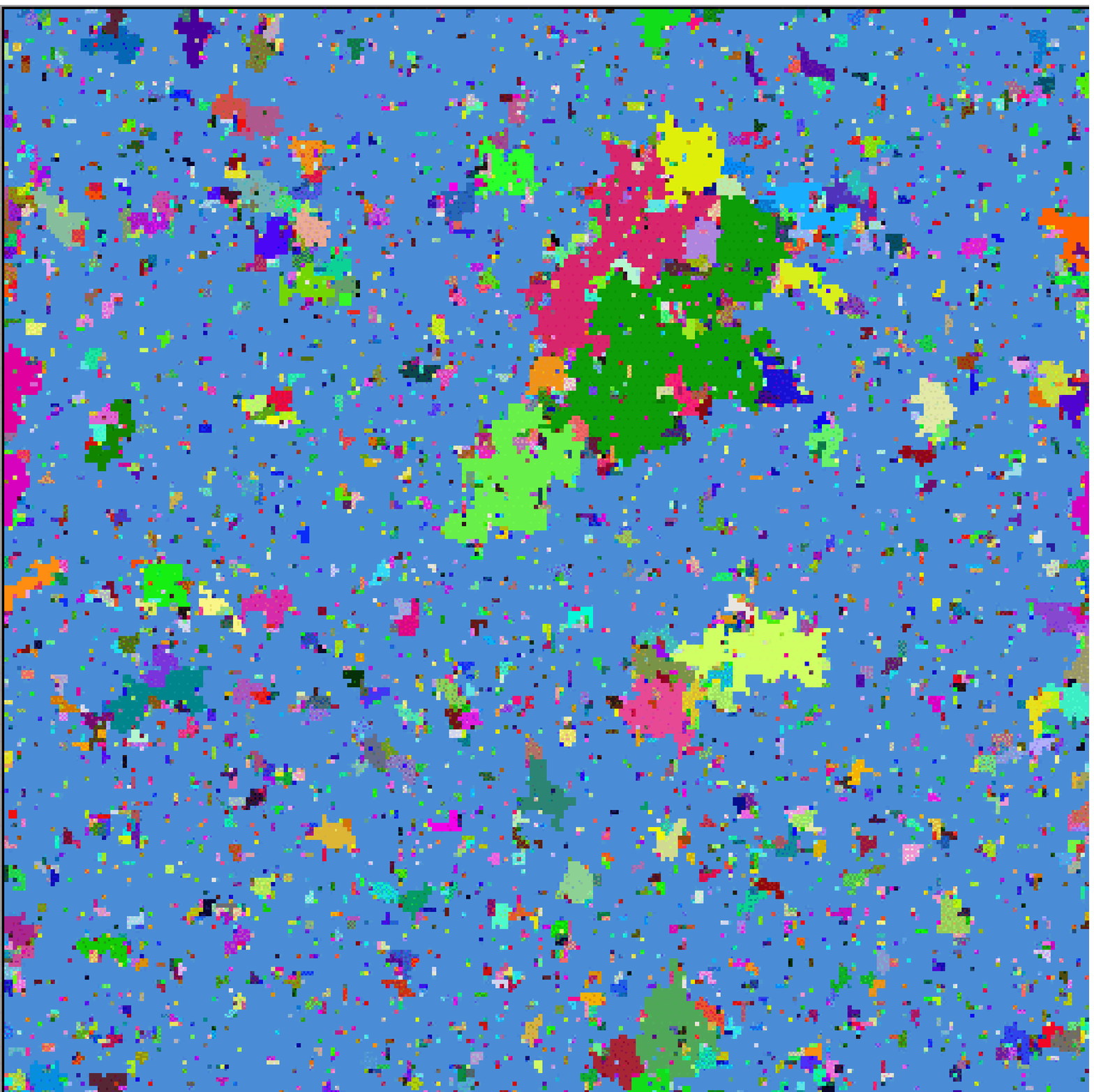}}

\centerline {
\includegraphics[width=5in]{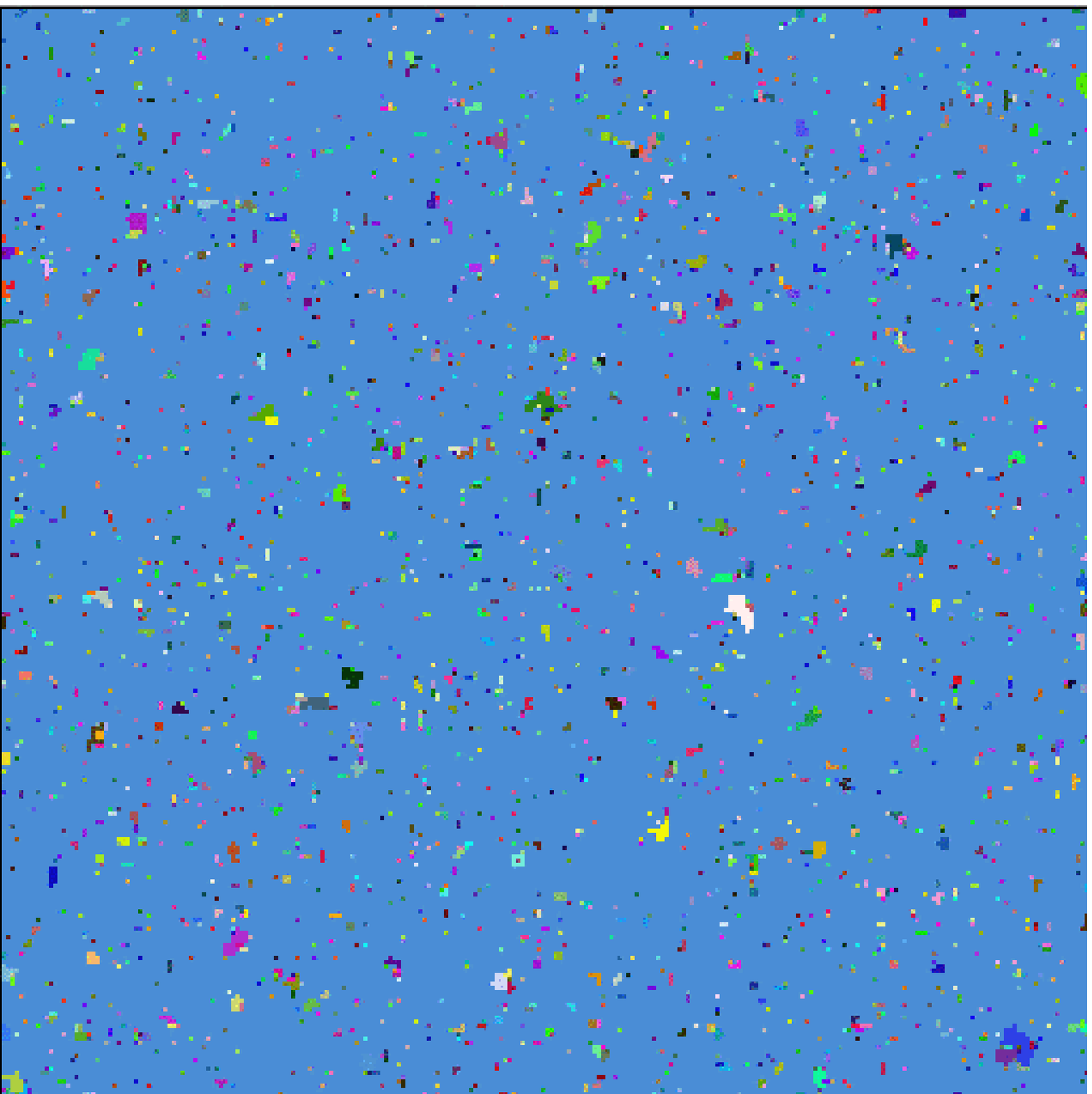}}

\section{Traditional percolation distribution}

What is the size distribution of clusters in percolation?  According to 
scaling theory (e.g., Stauffer \& Aharony 1994),
the number of clusters of size $s$ per site of the lattice behaves as 
\begin{equation}
n_s(p) = Êc_1 s^{-\tau} f(c_2 (p - p_c) s^\sigma)
\label{nsfscaling}
\end{equation}
where $\tau$ and $\sigma,$ are universal exponents, and $f(x)$ is a 
universal function.  However $c_1$ and $c_2$ are
non-universal, but vary with the lattice and percolation type being 
considered.  Here $f(0) = 1$ and $f(x) \to 0$ as $x \to \pm \infty$.
The scaling  relation can also be written as
\begin{equation}
n_s(p) = Êc_1 s^{-\tau} F_\pm(s/s^*)
\end{equation}
where $s^* = s_\xi$ is a typical size and scales as $|p - p_c|^{-1/\sigma}$.
If a ratio of moments such as $M_3/M_2$ is used to define
$s^*$, then $s^*$ will have a coefficient that
differs above and below $p_c$, and that difference will have to
be included in the definition of $F_\pm(z)$ in terms of $f(z)$.
In any case, $F_\pm(x)$
is  universal, but the non-universal coeffiecient $c_1$ remains.

Indeed, $n_s$ can never directly be put into a completely universal form because it
concerns the size $s$ which is a lattice-level quantity and by its very nature
non-universal.

\section{Universal form of the size distribution}
We have shown (Ziff, Lorenz, \& Kleban 1999) that, at the critical point $p_c$, the distribution
of cluster size can be written in a completely universal form:
\begin{equation}
N(\hbox{clusters whose enclosed area} \ge A) \sim \frac{C \Omega}{A} 
\label{NA}
\end{equation}
in a system of total area $\Omega$,
where $C$ is a universal constant which for 2d percolation is given by
(Cardy Ziff, 2003)
\begin{equation}
C = \frac{1}{8 \pi \sqrt 3}\approx 0.0229720373 
\end{equation}
Results have also been found for the more general Potts model.

This result can also be phrased:
\begin{itemize}
\item The number of clusters per unit area ${\cal A}$ whose enclosed area
is greater than ${\cal A}$  equals a universal number $C$.
\end{itemize}

\section{Proof of (5)}
From the definition of $n_s$ we have
\begin{equation}
N(\hbox{clusters whose size} \ge s) = \Omega \int_s^\infty n_s ds \sim  \Omega s^{1-\tau}
\end{equation}
Each cluster has one external hull which encloses an area 
\begin{equation}
A \sim r^2 \sim s^{2/D}
\end{equation}
where $D$ is the fractal dimension of
the cluster and $r$ is a length scale.   Thus, $s \sim A^{D/2}$ and 
\begin{eqnarray}
N(\hbox{clusters whose enclosed area} \ge A) &\sim&\Omega A^{(D/2)(1-\tau)} \nonumber \\
&=&{\Omega}/{A}
\label{NAproof}\\
\nonumber\end{eqnarray}
by virtue of the hyperscaling relation
\begin{equation}
\frac{d}{D} = \tau - 1
\end{equation}
Furthermore, the constant $C$ must be universal, since $N(\hbox{area} \ge A)$ is only a property of the larger clusters, or, in 
other words, a property of the  universal percolation {\em fractal}.

\section{Derivation of the exact  value of $C$}
Eq.\ (\ref{NA})
implies that the average {\em depth}, defined as the number of outer hulls crossed in going from
a point to the edge of the system, scales as 
\begin{equation}
C \log (\Omega/A_0)
\end{equation}
where $A_0$ is a lower
cutoff of the area ($\sim$ the lattice spacing).  Transforming conformally from an annulus
to a rectangle  using
$z \to \log z$, we find that this corresponds to a rectangular system with $8\pi C$ clusters crossing
per unit length.  But Cardy has shown that that number is $1/\sqrt{3}$, which implies
\begin{equation}
C = \frac{1}{8 \pi \sqrt 3} 
\end{equation}
as given above.

\section{Numerical demonstration}
We generate ``rooted" hulls by carrying out a hull generating
walk on a lattice, and stopping when the cluster closes or when an upper cutoff is reached.
Using a (virtual) lattice of $65536\times 65536$, and a cutoff of $2^{22} = 4194304$ steps, we
add a weight to each walk of a factor of 1/(no.\ steps), and thus find an {\em unbiased} 
estimate of $N($area $ = A)$ as long as $A < $cutoff.  We bin logarithmically,
\begin{equation}
N(A,2A) = N(\hbox{hulls w/enclosed area} \in(A,2A)) / \Omega
\end{equation}
which should behave asymptotically for large $A$ as 
\begin{equation}
N(A,2A) = C \left(\frac{1}{A} - \frac{1}{2A}\right) =  \frac{C}{2A}
\end{equation}
or 
$2 A N(A,2A) = C$ (or $2C$ for both interior and exterior hulls):

\centerline {
\includegraphics[width=5 in]{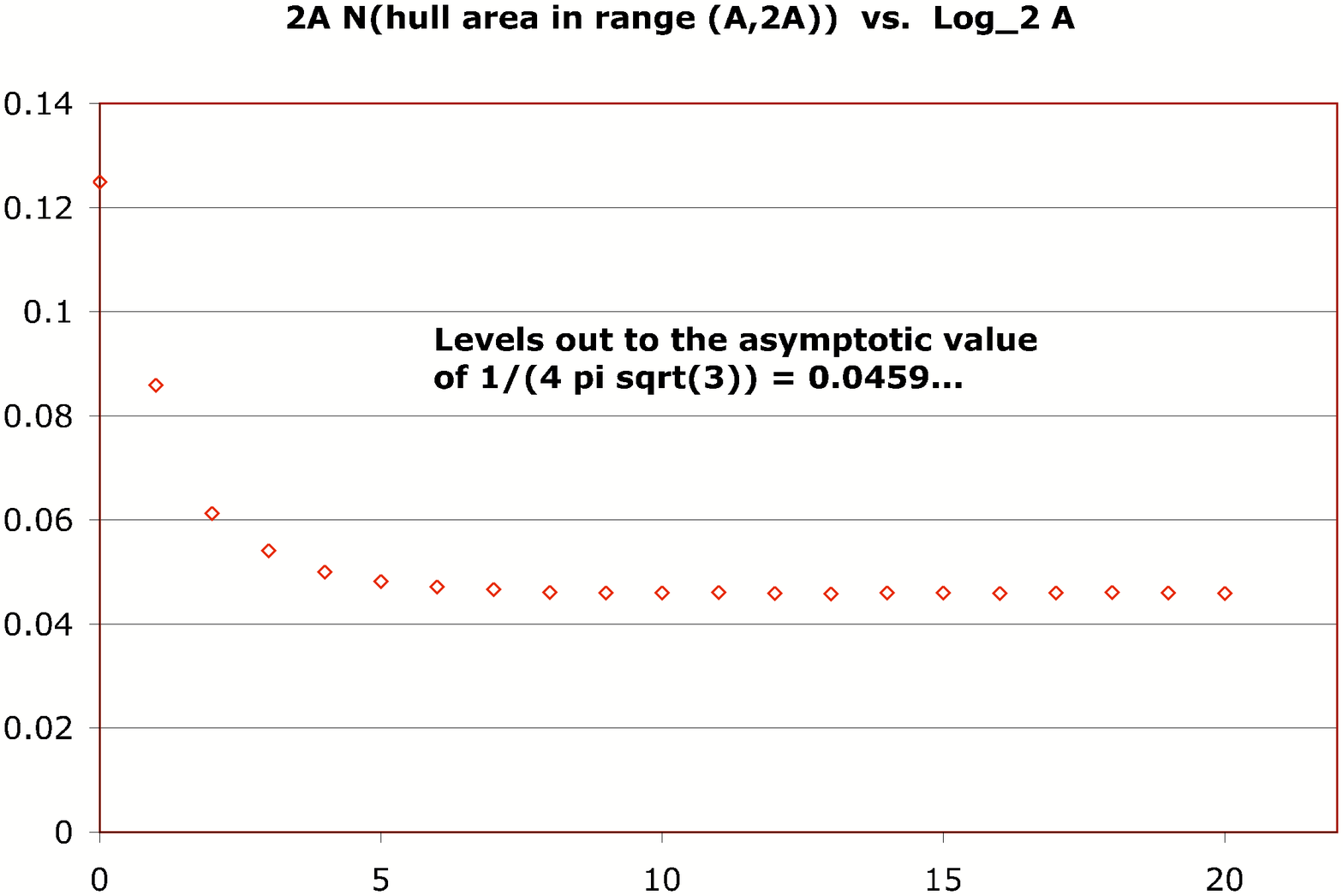}}

\section{Off-critical behavior}
 Away from $p_c$, we expect
\begin{equation}
N(\hbox{hulls whose enclosed area} \ge A) \sim \frac{C\Omega}{A} G(A/A^*)
\end{equation}
where $A^* =$  typical hull area and $G(x)$ is a 
scaling function that should be universal.   
One expects
\begin{equation}
A^* \sim \xi^2 \sim \frac{1}{|p - p_c|^{2\nu}}
\end{equation}
where $\xi$ is the correlation length and $\nu = 4/3$ in 2d,
since the area of a percolating cluster is not a fractal (even
though its boundary is a fractal).

Here we plot the same quantity as before but at $p = 0.495$.
The quantity $ A N(A, 2A)$ for both internal and external hulls
approaches the value of $2C$ for $\log A \approx 5$ but then
deviates for larger $A$.

\centerline {
\includegraphics[width=5 in]{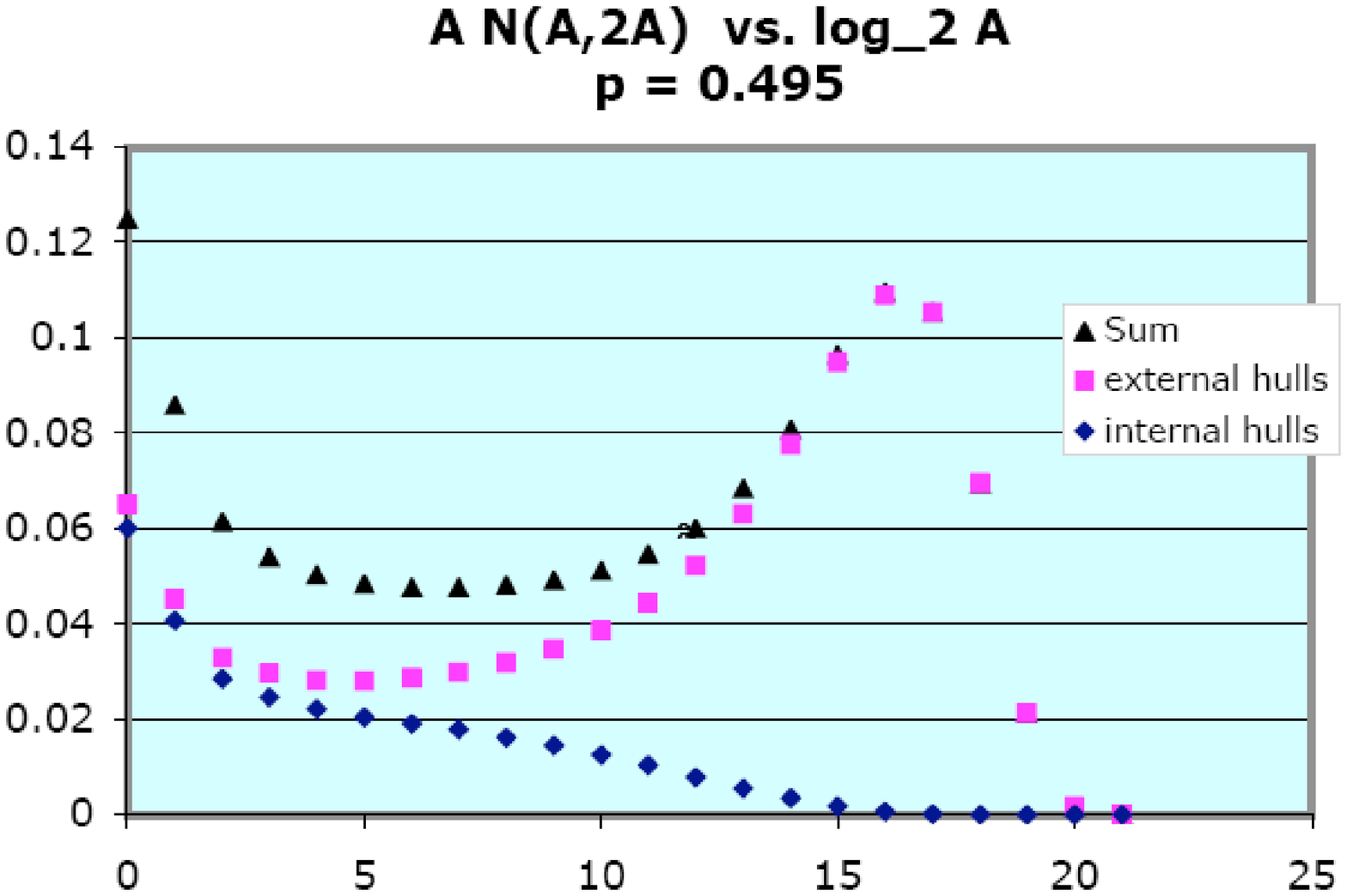}}

\section{Average Area}
We can define an average area by averaging $A^2$ over all 
clusters (normalizing by $\Omega$):

\begin{eqnarray}
\langle A \rangle &=& \frac{1}{\Omega} \sum_A A^2 N(\hbox{area } = A) \nonumber \\
&\approx & \frac{1}{\Omega} \int_0^\infty A^2  N(\hbox{area } = A) \, dA \nonumber \\
&=& \frac{1}{\Omega} \int_0^\infty 2A N(\hbox{area}\ge A)\,dA  \nonumber \\
&=& 2C \int_0^\infty   G(A/A^*)\, dA \nonumber \\
&=& 2C A^* \int_0^\infty   G(x)\, dx
\end{eqnarray}
so that 
\begin{equation}
\langle A \rangle \propto A^*
\end{equation}
Note that we do {\it not} use $\sum A\, N($area$ = A)$ to define
the average area, because that 
quantity scales as $\ln A^* / A_0$, where $A_0$ is a lower cutoff
(area of basic cell of the lattice).  In fact, that quantity
gives the
average {\it depth} of crossed loops.  
\section{Numerical measurement of average area}
We have verified that $\langle A \rangle$ indeed scales as $|p - p_c|^{-2\nu}$.

\begin{tabular}{lrr}
$p$&$\langle A \rangle$&logarithmic slope \\
\hline
$0.4 $&$    	39.786$&   \\
$0.42 $&$    	70.351$&$-2.554$\\
$0.44 $&$    	148.060$&$-2.587$\\
$0.455 $&$    	313.931$&$ -2.612$\\
$0.46 $&$    427.332$&$ -2.618$\\
$0.465 $&$    607.705$&$ -2.620$\\
$0.47 $&$    912.729$&$ -2.639$\\
$0.475 $&$    1478.133$&$ -2.628$\\
$0.48 $&$    2670.927$&$-2.646$\\
$0.485 $&$    5730.272$&$ -2.652$\\
$0.49 $&$    16841.970$&$ -2.659 $\\     
\end{tabular}

{\large Table. Average area size, per lattice site (both internal and external hulls) 
as a function of $p$.  The logarithmic slope $\Delta \log A / \Delta \log (p_c - p)$,
clearly approaches a value consistent with $-8/3$.}

\bigskip


Here it is plotted on a log-log scale:

\vspace{.2in}
\centerline {
\includegraphics[width=5 in]{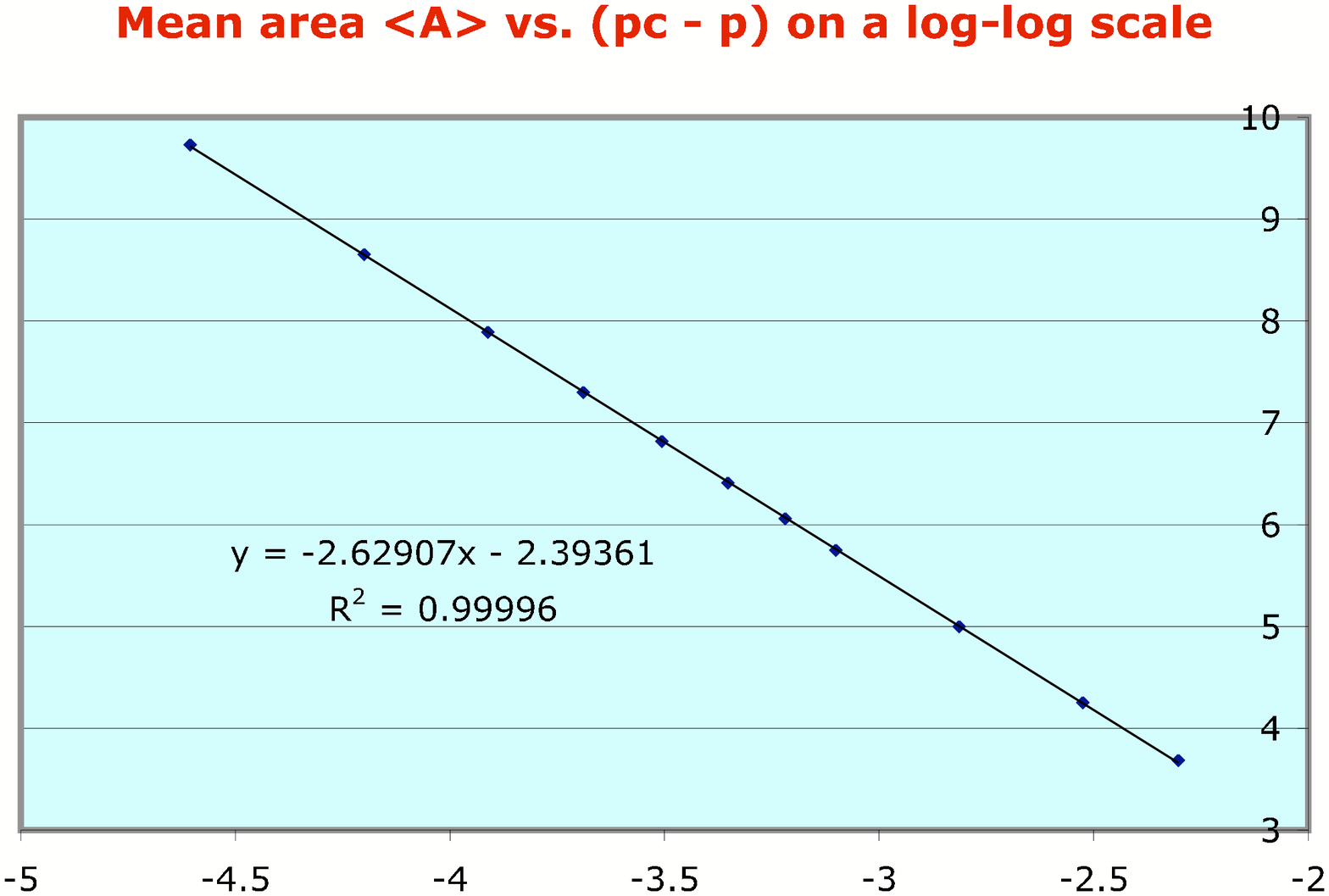}
}
\vspace{.2in}

\section{Higher-order behavior of the mean area}
Assuming an exponent of $8/3$, we can find the higher-order behavior
by plotting $(p_c-p)^{8/3} \langle A \rangle$ vs.\ $(p_c - p)^x$, and
varying $x$ until we find a straight line, which occurs for $x = 4/3$:

\centerline {
\includegraphics[width=5 in]{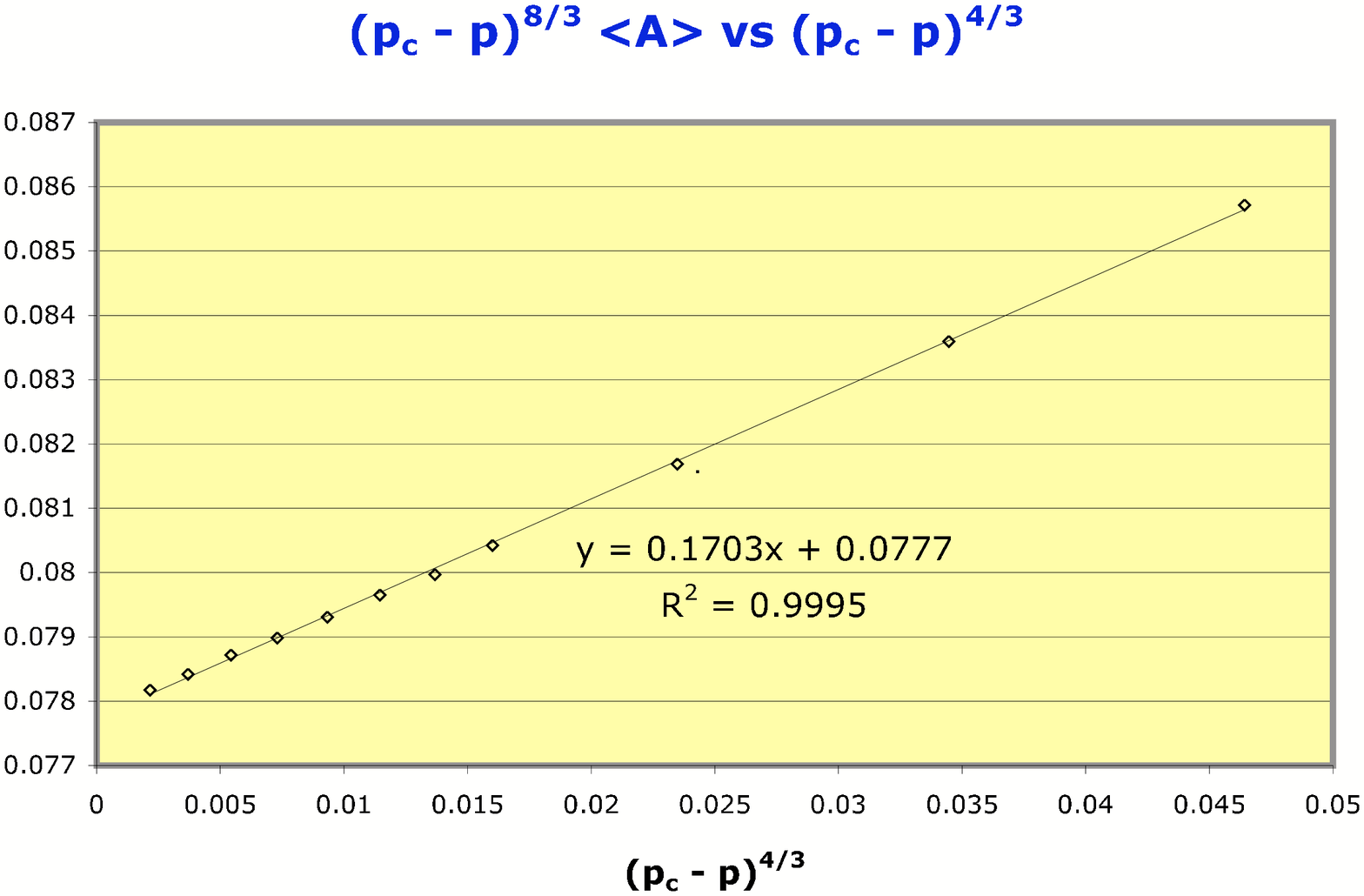}
}

which implies
\begin{equation}
\langle A \rangle = a_0 (p_c - p)^{-8/3} + a_1 (p_c - p)^{-4/3}\ldots
\end{equation}

\section{Asymptotic behavior of $G(x)$}

For large $x$, we find $G(x)$ decays as $e^{-cx}$:

\centerline {
\includegraphics[width=5 in]{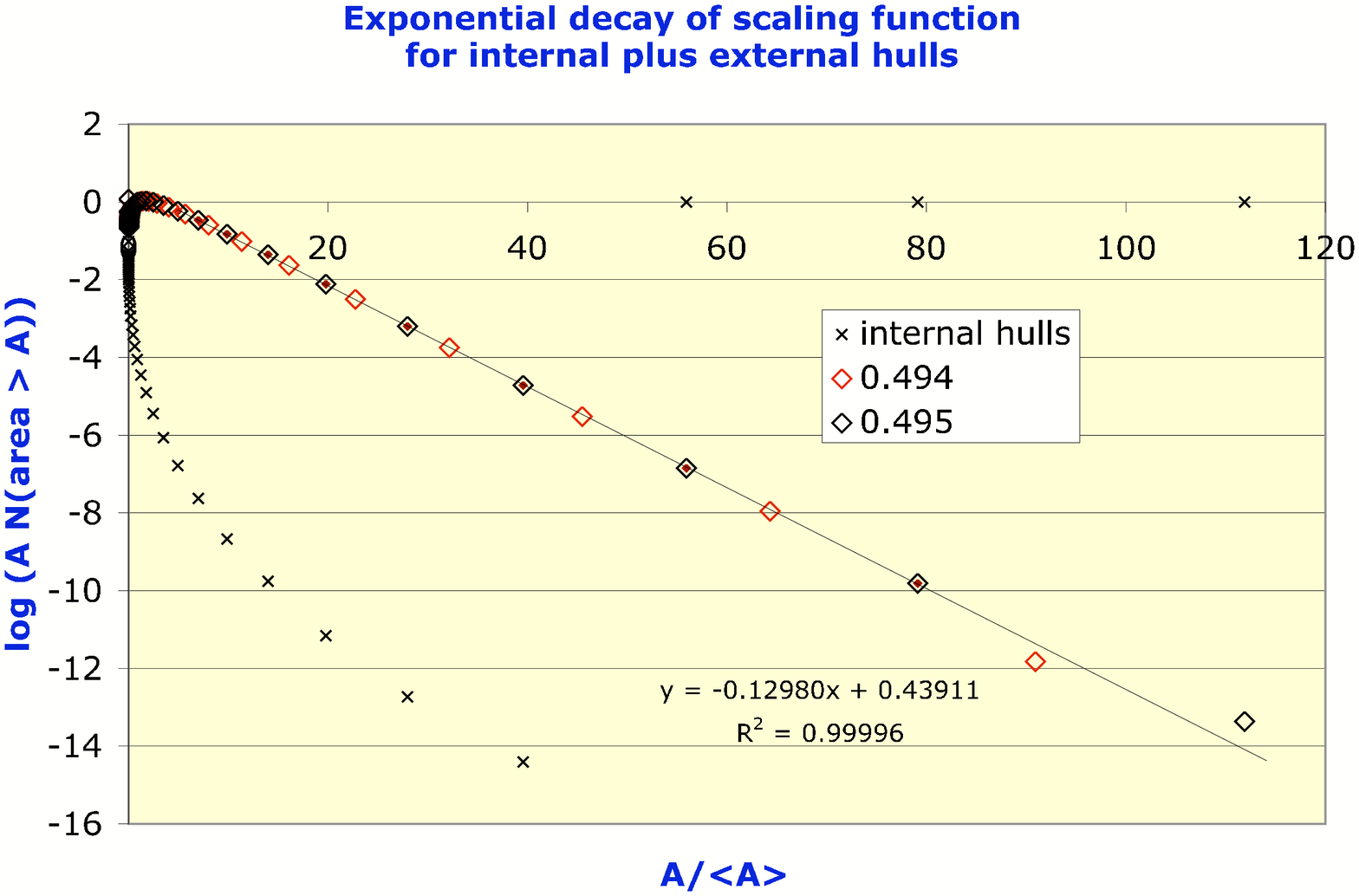}
}

Because the precise definition of $A^*$ is arbitrary up to a constant, we 
can fix that constant by requiring that $G(x)$ is a simple exponential for
large $x$:

\begin{equation}
G(x) \sim e^{-x}  \qquad x \gg 1 
\end{equation}
which implies that 
\begin{equation}
A^* = \langle A \rangle /0.130   
\end{equation}
so that
 \begin{equation}
2C  \int_0^\infty   G(x)\, dx = 0.130
\end{equation}

\section{Comparison to sub-critical percolation mass}
For $s \gg s^*$, the distribution of mass $s$ in subcritical
percolation is believed to behave as (Kunz and Souillard 1978):
\begin{equation}
n_s(p) \sim \frac{1}{s} \, e^{-as/s^*}
\end{equation} 
\begin{itemize}
\item  This is evidently consistent with our result only if $s \sim A$
(instead of $s \sim A^{D/2} = A^{91/96}$) for subcritical clusters 

\item  Note also that $n_s$ apparently has a more complicated
form than $N$, since $n_s$
changes its leading power term from $s^{-\tau}$ for $s \ll s^*$,
to $s^{-1}$ for $s \gg s^*$, while $N$ always has the same
leading behavior of $A^{-1}$ for all $A$. 
\end{itemize}

\section{Amplitude ratio of average areas}

According to the universal of $G(x)$, the amplitude ratio of 
\begin{equation}
 \frac{\langle A \rangle^{\rm [ext]}} {\langle A \rangle^{\rm [int]}}
 =
 \frac{\int_0^\infty   G(x)^{\rm [ext]}\, dx}
{ \int_0^\infty   G(x)^{\rm [int]}\, dx} 
\end{equation}
is also universal. (Note -- there is only one $A^*$ in the system, used
for scaling for both internal and external clusters).
The measurements suggest a value of about 180 for this ratio
(note higher statistical error as $p \to p_c$):

\centerline {
\includegraphics[width=5 in]{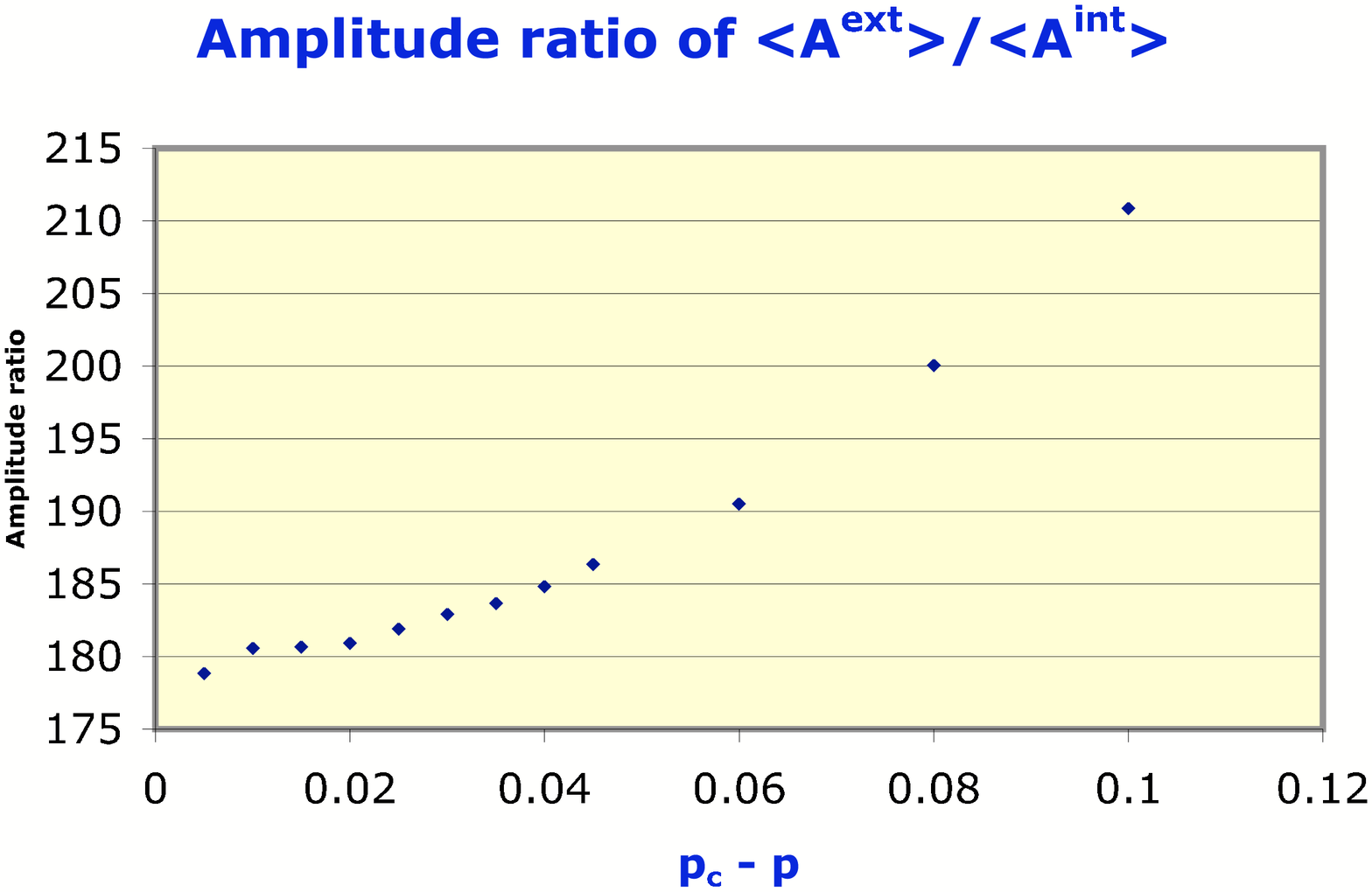}
}

\section{Alternate form of scaling}
We can also write $N$ as
\begin{equation}
N(\hbox{area} > A)
  = \frac{C \Omega}{A}g(\pm (A/A^*)^{1/2\nu})
\end{equation}
where $1/2\nu = 3/8$ and $\pm$ refers to external ($+$)
and internal hulls ($-$).  We write the argument of $g$
in  this way because
then it is proportional to $(p - p_c)A^{1/2\nu}$
and therefore $g(z)$ should be a single analytic function
of $z$ for both positive and negative $z$, as in (\ref{nsfscaling}) for the
size distribution.  
Here are our results for $p = 0.495$ for the scaling function
$g(x)$:

\centerline {
\includegraphics[width=4in]{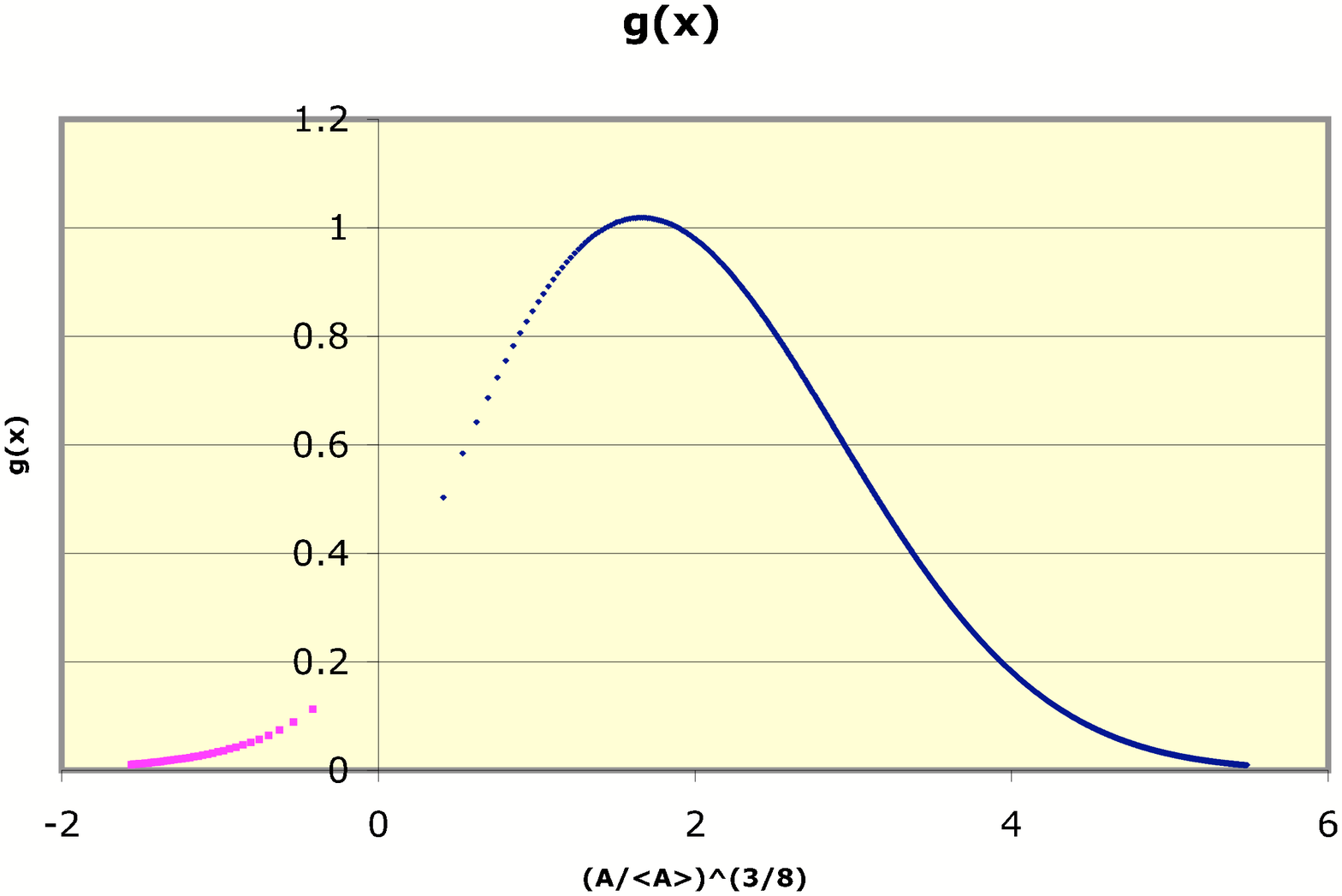}
\label{g}
}

\section{Conclusions}
We have shown
\begin{equation}
N(\hbox{hulls whose enclosed area} \ge A) \sim \frac{C\Omega}{A} G(A/A^*)
\end{equation}
with $G(x)$ a unique universal function satisfying
\begin{equation}
G(x) = 1  \qquad x = 0 
\end{equation}
\begin{equation}
G(x) \sim c e^{-x}  \qquad x \gg 1
\end{equation}
where  $c \approx exp(0.439) = 1.55$ and $A^*=  \langle A \rangle / 0.130$.

\section{Appendix: Scaling for N(area = A)}
For $N_A = $ number of hulls whose area {\em equals} $A$,
we can write
\begin{equation}
N_A = \frac{C}{A^2} \overline F_\pm(A/A^*)
  = \frac{C}{A^2}\overline f(c_3 (p-p_c) A^{1/2\nu})
\end{equation}
where $1/2\nu = 3/8$ and $\overline f(x) = \overline F(x^{2\nu})$.
This is not a universal form of the
distribution both because it involves clusters of {\em exactly}
area $A$, and also because the non-universal constant
$c_3$ shows up in the argument of $\overline f$ (in the second
form).  It is interesting
however as the counterpoint of the scaling form (\ref{nsfscaling}) of
the percolation mass that is also analytic at $p = 0$. 

Here we keep internal and external hulls separate.  
By duality, we have the symmetry, 
\begin{equation}
\overline f^{[\rm{ext}]}(x) =\overline f^{[{\rm int}]}(-x)
\end{equation}
which allows us to  
the entire curve of $\overline f(x)$ by making measurements
at $p < p_c$ only, and using the internal hulls for $-x$
and the external hulls for $x$.
Here are the results
for $p = 0.44$, where we plot $A^2 N_A$ vs.\  $(p-p_c) A^{3/8})$,
analogous to (\ref{g}):

\centerline {
\includegraphics[width=5 in]{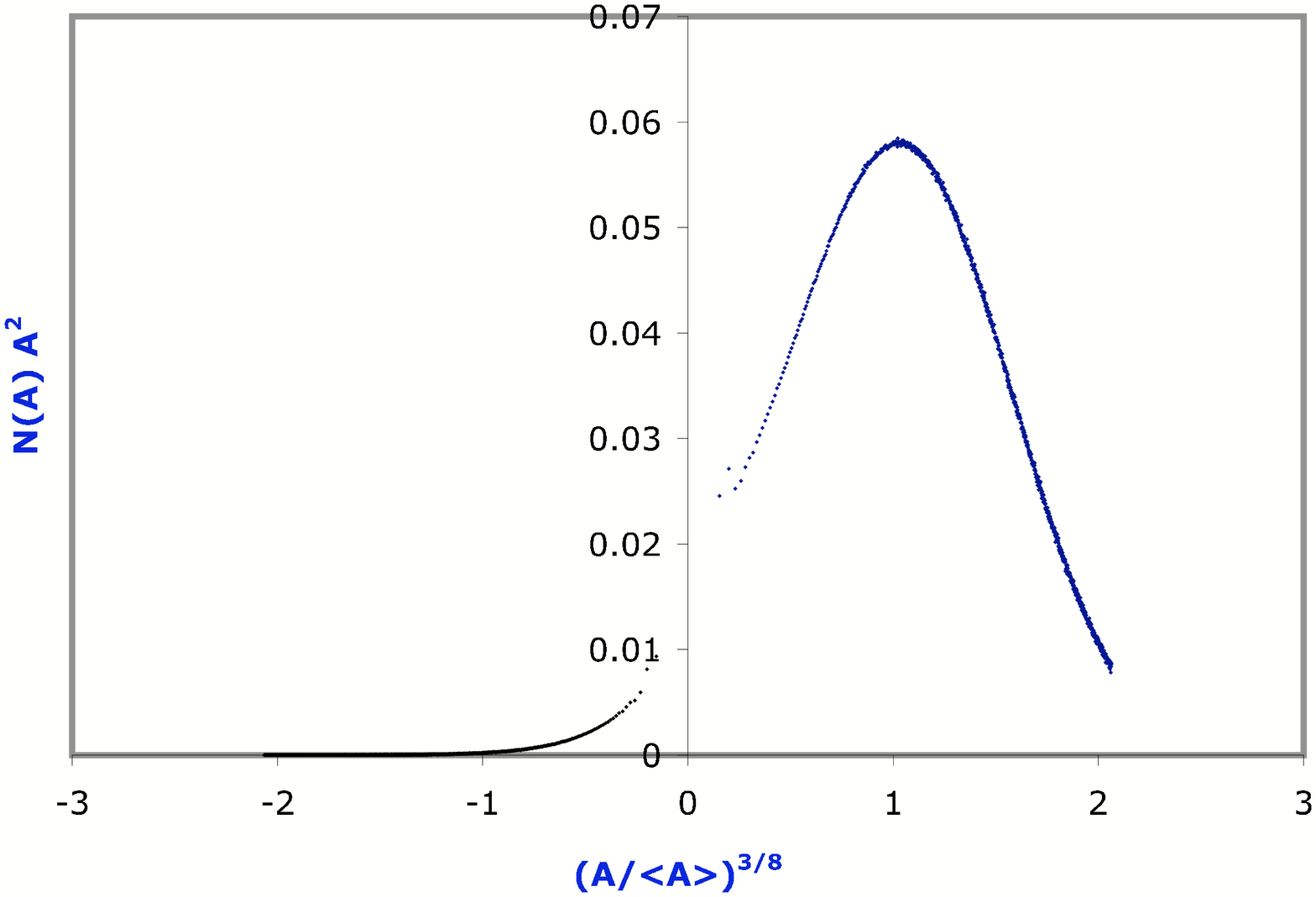}
}

\section{Note}
An inconsistent factor of 4 was used in
the definition of the area, and this error
may not have been consistently corrected in 
all of the plots.
}

\section{References}
\smallskip
\frenchspacing

\noindent 1. D.~Stauffer and A.~Aharony, {\em Introduction to Percolation Theory, 2nd. edition} (Taylor \& Francis, London, 1994).
\smallskip

\noindent 2. M. E. J. Newman and R. M. Ziff, 
Efficient Monte Carlo algorithm and high-precision results
for percolation, Phys. Rev. Lett. {\bf 85} 4104-4107 (2000),
and Fast Monte Carlo algorithm for site or bond percolation,
Phys. Rev. E {\bf 64}, 016706 (2001).

\smallskip
\noindent 3. J. Cardy and R. M. Ziff, Exact results for the universal
area distribution in percolation, ising, and Potts models,
J.\ Stat.\ Phys.\ {\bf 110}, 1-33 (2003).

\smallskip
\noindent 4. R. M. Ziff, C. D. Lorenz, and P. Kleban,
Shape-dependent universality in percolation,
Physica A {\bf 266}, 17-26 (1999).

\smallskip

\noindent 5. H. Kunz and B. Souillard, Essential singularity
in the percolation model, Phys. Rev. Lett. {\bf 40}, 133-135 (1978).
\nonfrenchspacing

 \end{document}